\documentclass{article}
\pdfoutput=1
\usepackage{jcappub}

\def\lbldef#1#2{\expandafter\gdef\csname #1\endcsname {#2}}

\def\href#1#2{#2}

\usepackage{graphicx}

\newcommand{\prob}{{\rm Pr}}
\newcommand{\data}{{\mathbf d}}
\newcommand{\pars}{{\boldsymbol \alpha}}
\newcommand{\hnought}{{H_0}}
\newcommand{\neff}{{N_{\rm eff}}}
\newcommand{\yhe}{{Y_{\rm He}}}
\newcommand{\fnu}{{f_{\nu}}}
\newcommand{\mnu}{{M_{\nu}}}
\newcommand{\sigeight}{{\sigma_8}}
\newcommand{\nrun}{{\alpha}}
\newcommand{\omgmh}{{\Omega_c h^2}}
\newcommand{\omgnuh}{{\Omega_\nu h^2}}
\newcommand{\zeq}{{z_{\rm eq}}}

\newcommand{\be}{\begin{equation}}
\newcommand{\ee}{\end{equation}}
\newcommand{\bea}{\begin{eqnarray}}
\newcommand{\eea}{\end{eqnarray}}
\newcommand{\bc}{\begin{center}}
\newcommand{\ec}{\end{center}}

\title{Is there evidence for additional neutrino species from cosmology?} 

\author[a]{Stephen M. Feeney,}
\author[a]{Hiranya V. Peiris}
\author[b,c]{and Licia Verde}

\emailAdd{stephen.feeney.09@ucl.ac.uk}
\emailAdd{h.peiris@ucl.ac.uk}
\emailAdd{liciaverde@icc.ub.edu}

\affiliation[a]{Department of Physics and Astronomy, University College London, London WC1E 6BT, U.K.}
\affiliation[b]{ICREA \& ICC, Institut de Ciencies del Cosmos, Universitat de Barcelona (IEEC-UB), Marti i Franques 1, Barcelona 08028, Spain}
\affiliation[c]{Theory Group, Physics Department, CERN, CH-1211, Geneva 23, Switzerland}

\abstract{It has been suggested that recent cosmological and flavor-oscillation data favor the existence of additional neutrino species beyond the three predicted by the Standard Model of particle physics. We apply Bayesian model selection to determine whether there is indeed any evidence from current cosmological datasets for the standard cosmological model to be extended to include additional neutrino flavors. The datasets employed include cosmic microwave background temperature, polarization and lensing power spectra, and measurements of the baryon acoustic oscillation scale and the Hubble constant. We also consider other extensions to the standard neutrino model, such as massive neutrinos, and possible degeneracies with other cosmological parameters. The Bayesian evidence indicates that current cosmological data do not require any non-standard neutrino properties.}

\begin{document}

\maketitle


\section{Introduction}
\label{sec:intro}

In the past decade there has been great progress in neutrino physics: in particular, significant evidence has been found for non-zero neutrino masses from observations of neutrino oscillations. The Standard Model of particle physics has three massless neutrinos, and thus massive neutrinos require beyond-the-Standard-Model physics or uncomfortable fine-tuning. When considering extensions of the Standard Model, it is not unreasonable  to explore the possibility of more than three neutrino species. This subject has recently received renewed attention (see Ref.~\cite{Abazajian:2012ys} and references therein for a thorough review): in particular, extra sterile neutrinos have been invoked to explain  recent anomalies observed in neutrino experiments, most notably results from the LSND and MiniBooNe experiments \cite{miniboone, miniboone2, miniboone3, giunti}.

The expansion history of the early Universe depends on the physical energy density in relativistic particles, which can be expressed in terms of the energy density in photons (highly constrained by the measurement of the cosmic microwave background (CMB) temperature) and an effective number of neutrino species $N_{\rm eff}$.\footnote{In the Standard Model $N_{\rm eff}=3.046$, differing from the number of neutrino species $N_{\nu}=3$ to account for QED effects, for neutrinos being not completely decoupled during electron-positron annihilation  and other small effects, see e.g. Refs.~\cite{lesgourguespastor,mangano}.} Cosmology therefore offers two windows to constrain $N_{\rm eff}$: nucleosynthesis, where the abundances of the light elements are driven by the Hubble rate, and large-scale structures -- in particular the CMB -- which probe the radiation density at matter-radiation equality and at recombination.

A deviation from the Standard Model prediction for $N_{\rm eff}$ is not necessarily evidence for new neutrino physics: any unorthodox early-Universe expansion history, for example due to the presence of non-standard energy density (dubbed ``dark radiation"), can be parametrized as a deviation from the expected number of neutrino species.\footnote{Provided the extra radiation responsible for changing the expansion is free-streaming: interacting dark radiation (and, indeed, neutrinos) impart different signatures (see Refs.~\cite{Bash_and_Seljak:2004,Bell_etal:2006,Cyr-Racine_and_Sigurdson:2012}).} Limits on possible deviations from $\neff = 3.046$ therefore place constraints on neutrino physics and on any other process that changes  the expansion history~\cite{Lopez:1998jt,Hannestad:1998zw,Kaplinghat:2000jj,Trotta_and_Melchiorri:2005,Davoudiasl:2007jr,Calabrese:2012vf,GonzalezGarcia:2010un,Lizarraga_etal:2012}. In this work we concentrate on constraints from the CMB and large-scale structure, phrasing the results in terms of neutrinos rather than dark radiation. Current cosmological evidence for extra neutrinos from CMB and large-scale structure observations is discussed in Refs.~\cite{Hamann:2010bk,Calabrese:2011hg,Giusarma:2011ex, Archidiacono:2011gq, Giusarma:2011zq, Archidiacono:2012gv, Archidiacono:2012ri,Hou:2011ec,Abazajian:2012ys,Joudaki_etal:2012,Giusarma_etal:2012,Zhao_etal:2012,R-S_etal:2013,R-S_etal:2013b}; however, see Refs.~\cite{Hamann_etal:2007,reid,GonzalezMorales:2011ty} for a different interpretation. 

Given the fact that the Standard Model firmly predicts three neutrino families, and that dark radiation would signal that the standard cosmological model, $\Lambda$CDM, is incorrect, the issue of whether cosmology favors extra neutrino species is an issue of {\em model selection} rather than parameter estimation. In model selection, which must be performed within the Bayesian framework to be self-consistent~\cite{Cox:1946}, the statistic used to select between models is the {\em Evidence}, $E$. The Evidence is the ``model-averaged likelihood" -- i.e. the integral of the likelihood over the parameter volume, weighted by the prior -- and thus encodes the full predictive power of the model. The Evidence has arbitrary normalization and is thus not useful in its own right; however, the ratio of the Evidence values for two models given the same data expresses the relative odds that these models are responsible for the observed state of the Universe. We capitalize the Bayesian Evidence where used to help distinguish it from the colloquial ``evidence".

While neutrinos have mass, the standard cosmological model traditionally contains massless neutrinos under the assumption that neutrino masses are too small to leave a measurable effect on cosmological observables. However, forthcoming cosmological data will have enough statistical power to detect the signature of non-zero neutrino masses, even if they are close to the lower limit allowed by oscillation experiments~\cite{core:2011,euclid:2011}; they have the potential to constrain the sum of the neutrino masses, $\mnu = \sum_1^{N_\nu} m_\nu$. We therefore also consider the Evidence for models with massive neutrinos, as well as possible degeneracies of neutrino properties with other cosmological parameters. Recent constraints on $\mnu$ from a range of cosmological datasets include Refs.~\cite{Mantz_etal:2010,Thomas_etal:2010,Komatsu_etal:2011,vanEngelen:2012va}.

The paper is organized as follows. In Sec.~\ref{sec:method} we review the Bayesian Evidence, its interpretation, calculation and alternatives, and introduce the datasets and models we explore. In Sec.~\ref{sec:results} we present our results. We compute standard Bayesian parameter estimates and prior-independent profile likelihood ratios in addition to Evidence ratios, which are presented for different model-dataset combinations. We are therefore able to compare our model-selection results to the parameter estimates traditionally used in the literature, as well as investigating the effects of the choice of priors on our findings. Finally, we conclude in Sec.~\ref{sec:conclusions}.


\section{Method and Datasets}
\label{sec:method}

The majority of neutrino analyses to-date employ parameter estimation to assess the need for additional neutrino physics, using different combinations of datasets to determine the most likely values of $\neff$ and $\mnu$. These analyses are typically carried out in the Bayesian framework (see Ref.~\cite{Abazajian:2012ys} for a review) but there are exceptions, such as Refs.~\cite{reid,GonzalezMorales:2011ty}. While such analyses provide estimates of the most likely parameter values given a particular model of neutrino physics, they are unable to provide a comparison of the relative probabilities of different models.

A deviation from the standard value of $\neff$ would signify that one or both of the standard models of cosmology and particle physics are incorrect or incomplete. Extended models, in which the neutrino mass and/or number of species are allowed to vary from the standard values, should therefore be compared with the standard model using model selection. To perform model selection, we must work within the Bayesian framework,  in which models are selected based upon their relative Evidence values; parameter constraints are also produced as a by-product of the Evidence calculation. To explore the prior-dependence of our results, we also report a frequentist approximation to model selection -- the profile likelihood ratio -- which relies only on the likelihood and is thus prior-independent.

Our work is novel not only because we consider the Bayesian Evidence for additional neutrino physics, but also because we include measurements of weak gravitational lensing of the CMB by large-scale structure (see e.g. Ref.~\cite{Hanson_etal:2010} for a review) in our analysis. The CMB lensing signal, first detected by the Atacama Cosmology~\cite{Das:2011ak} and South Pole~\cite{vanEngelen:2012va} Telescopes, peaks at redshifts of $\sim 2$ and is sensitive to the growth of structure. Adding the CMB lensing power spectrum to the temperature power spectrum therefore adds information about the late-time Universe, including the effect of massive neutrinos.


\subsection{Model Selection and Bayesian Evidence}

The Bayesian Evidence $E$ for a model $M$ describing a data vector $\data$ with an $N$-dimensional set of parameters $\pars$ is given by
\begin{equation}
E=\int {\rm d} \pars \, \prob(\data | \pars, M) \, \prob(\pars | M),
\label{eq:evidence}
\end{equation}
where $\prob(\data | \pars, M)$ is the likelihood (the probability of obtaining the data given the model and a particular set of parameter values) and $\prob(\pars | M)$ is the prior probability on the model parameters. Eq.~\ref{eq:evidence} is a multi-dimensional integral over a volume defined by the parameter ranges permitted by the prior: it is the probability of the obtaining the data given the full model, $\prob(\data|M)$, rather than given a particular set of parameters. The Evidence depends both on the height of the likelihood and the total volume of the parameter space: thus  a model in which the likelihood is significantly non-zero over a large fraction of the permitted parameter volume is more predictive than a model where a small perturbation from the best-fit parameters yields a significant degradation in fit.

The Evidence describes the probability of getting a set of measurements given a model; for model selection, we want to instead compare the probabilities of a pair of models being true given a set of data. Bayes' theorem~\cite{Bayes_Price:1763} allows us to relate these quantities as
\begin{equation}
\frac{\prob(M_1|\data)}{\prob(M_2|\data)}=\frac{\prob(M_1)}{\prob(M_2)}\frac{E_1}{E_2}\,,
\end{equation}
where the {\em a priori} preference for model $M_1$ over $M_2$, $\prob(M_1)/\prob(M_2)$, is typically set to unity. Under this assumption, the Evidence ratio, $E_1/E_2$, between two models yields the ratio of probabilities for the two models in the light of the data, or equivalently the relative ``betting odds'' for two models.

Evidence ratios are often compared using the Jeffreys' scale~\cite{Jeffrey} (in fact, a slightly modified version~\cite{kassraftery}), which rates $| \Delta \ln E|  = | \ln (E_1/E_2) |< 1$ as being ``not worth a bare mention'', whereas $| \Delta \ln E | > 5$ is regarded as ``highly significant''. Values of $1< | \Delta \ln E| <2.5$ indicate ``substantial'' evidence for the model with higher $E$; values in the range $2.5< | \Delta \ln E| <5$ indicate ``strong" evidence. In terms of odds, ``substantial'' evidence corresponds roughly to 12:1; ``strong'' evidence to 150:1. It is important to note that this scale is empirically calibrated and should be used only as a guide: strict adherence to it carries all of the risks inherent in the use of any thresholded scale.

We use the publicly available {\tt camb}~\cite{camb} Boltzmann solver code to compute the theoretical CMB angular power spectra and BAO scale for different cosmologies, combined with a {\tt MultiNest}-enabled~\cite{Feroz:2008xx, Feroz:2007kg} version of the {\tt CosmoMC}~\cite{cosmomc} package to compute the Bayesian Evidence and parameter estimates. We use 800 {\tt MultiNest} live points and set the tolerance and efficiency parameters to their recommended values of 0.3 to ensure the accuracy and precision of the Evidence calculation. With these settings, the precision of the Evidence calculation is $\sigma_{\ln E} \sim 0.3$.

The Evidence is, by definition, sensitive to the prior choice. It is therefore extremely important to set the priors independently of the data used and, in some cases, to explore the prior dependence. We pay 
careful attention to the prior ranges in Section~\ref{sec:priors}.


\subsection{Profile Likelihood Ratio}

The Bayesian Evidence provides a self-consistent framework with which to perform model selection, but, as with all Bayesian methods, it depends on the prior choice. In some cases the priors can be physically motivated, but when this is not possible the Bayesian answer to the model selection issue cannot be considered definitive. It is therefore interesting to consider statistics which rely only on the likelihood and are thus prior-independent, even if these do not provide a fully consistent model-selection criterion. Here we use the profile likelihood ratio (PLR) (see e.g. Refs.~\cite{PLR, reid, GonzalezMorales:2011ty}). Assuming a model, $M$, has $N-1$ uninteresting parameters $\pars$ (which are marginalized over in the standard Bayesian approach) and one parameter, $\beta$, on which we want to report constraints, the PLR is the ratio between the conditional maximum likelihood for a fixed value of $\beta$ (say, $\beta_*$) and the unconditional maximum likelihood:
\begin{equation}
{\rm PLR}(\beta_*) = \frac{ \max \left[ \prob(\data | \pars, \beta = \beta_*, M) \right] }{ \max \left[ \prob(\data | \pars, \beta, M) \right] } \, .
\label{eq:plr_theory}
\end{equation}
In other words, the PLR is the ratio of the maximum likelihood for each value of $\beta$ to the overall maximum likelihood (allowing the other parameters to take {\em any} permitted values).

This quantity has an interpretation similar to the $\Delta \chi^2$, where the effective chi-square is identified with $-2\ln \prob(\data | \pars, \beta, M)$ and, by construction, is prior-independent. Assuming Gaussian statistics, one may therefore associate the PLR values of 0.5 and 2 with one- and two-$\sigma$ confidence limits. The PLR is particularly useful in testing nested models: cases where a more-complex model has a free parameter ($\beta$) which is fixed in the simpler model. The extensions to $\Lambda$CDM considered in this work fall into this category. If the best-fit value for this parameter differs from the standard, nested value at $> n\sigma$ one reports an $n$-$\sigma$ ``evidence'' for that parameter, and thus for the associated model. One must keep in mind that the confidence intervals may not have strict frequentist coverage, especially if the likelihood is far from Gaussian; nevertheless it allows us to investigate whether preferences for extra parameters in posterior confidence intervals are driven by the data or by the prior.

Eq.~\ref{eq:plr_theory} describes the ideal profile likelihood ratio, which would be calculable given an infinitely-fine sampling regime in the parameter $\beta$. In practice, given the sampling frequencies accessible to nested sampling or MCMC methods, one must instead calculate a proxy to the PLR, where the complex model's likelihood is first binned in the parameter of interest. The PLR is then given by the ratio of these binned maximum likelihoods to the maximum likelihood overall:
\begin{equation}
{\rm PLR}(\beta_*) \simeq \frac{ \max \left[ \prob(\data | \pars, \beta =  \beta_* \pm \Delta\beta, M) \right] }{ \max \left[ \prob(\data | \pars, \beta, M) \right] } \, .
\label{eq:plr_proxy}
\end{equation}


\subsection{Datasets}

\begin{figure}[tb]
\centering
\includegraphics[width=7.5cm]{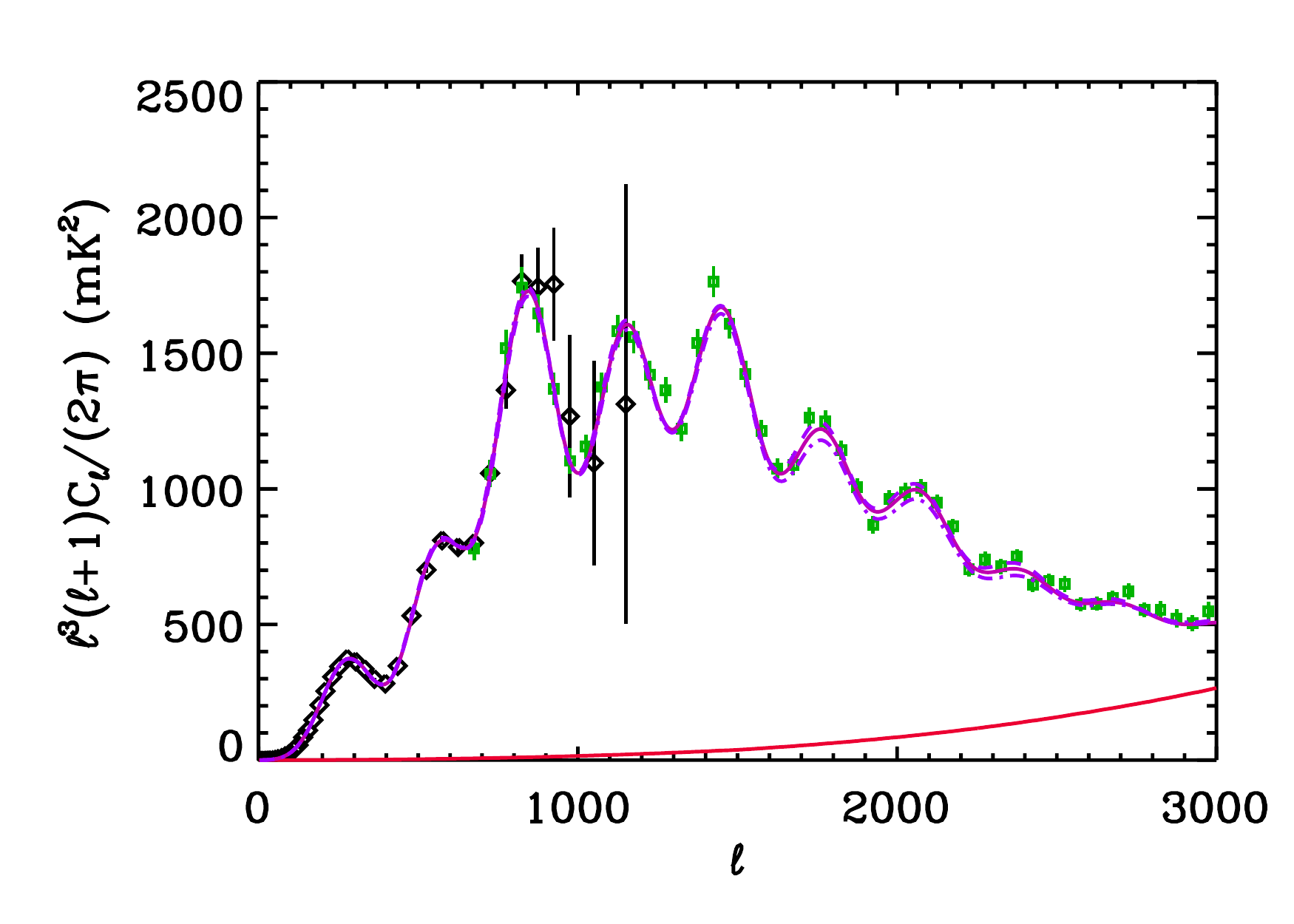}
\includegraphics[width=7.5cm]{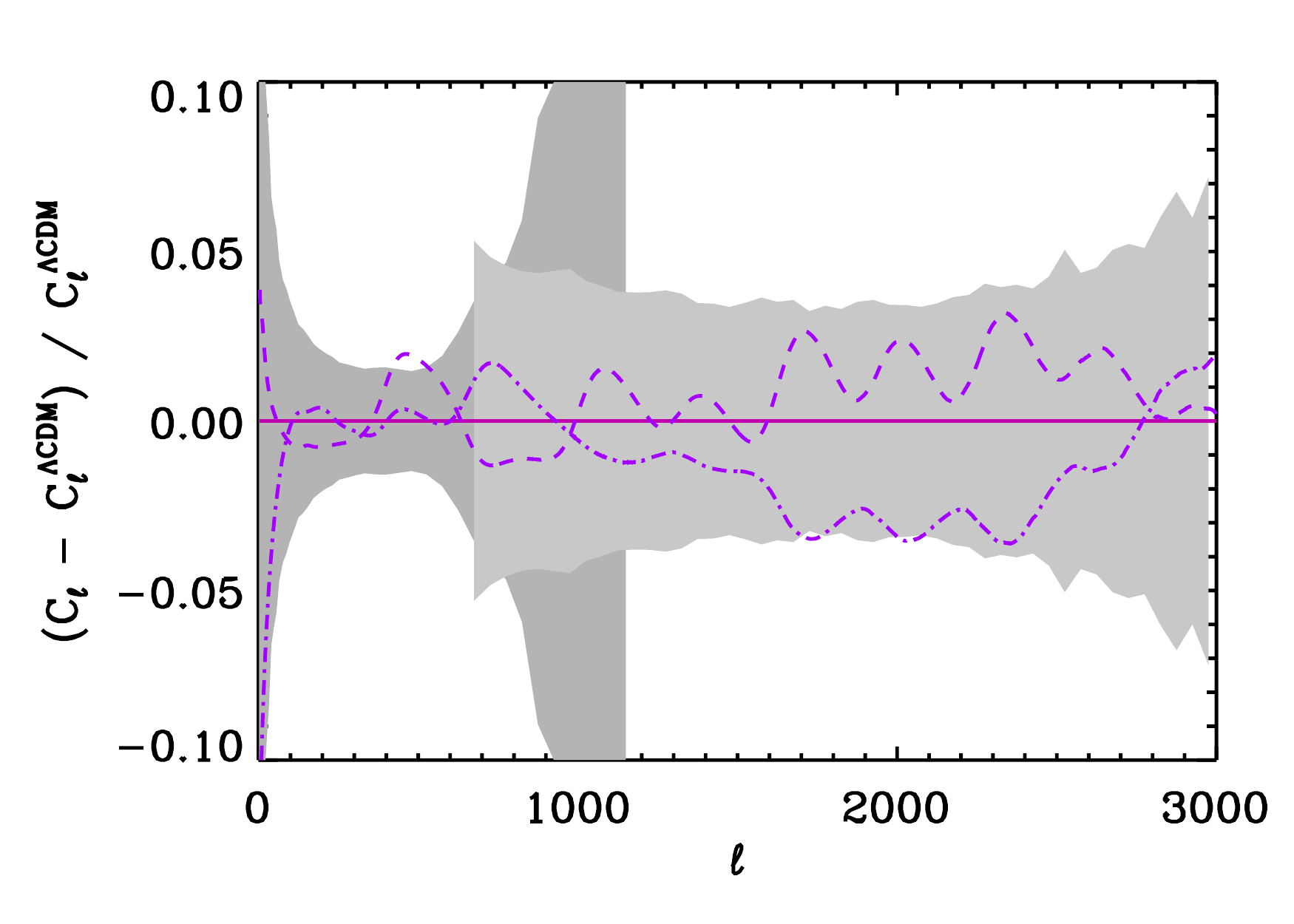}
\caption{Left: CMB temperature power spectrum measurements from the WMAP satellite (black diamonds) and the South Pole Telescope (green squares). The SPT data include foreground contributions from Poisson-distributed and clustered point sources, as well as the kinetic and thermal Sunyaev-Zel'dovich effects; the scale of this contribution is indicated by the red solid line. To illustrate the degeneracy between $\neff$ and $\omgmh$, power spectra for nearly-degenerate models with $\neff$ equal to $3.046$ (solid), $2.0$ (dashed) and $5.0$ (dot-dashed) are overlaid on the WMAP and SPT data. Right: the fractional differences of the power spectra with respect to $\Lambda$CDM are shown as a function of scale, with error bars from WMAP and SPT indicated by the dark- and light-grey regions, respectively.}
\label{fig:CMBCl}
\end{figure}

\begin{figure}[tb]
\centering
\includegraphics[width=7.5cm]{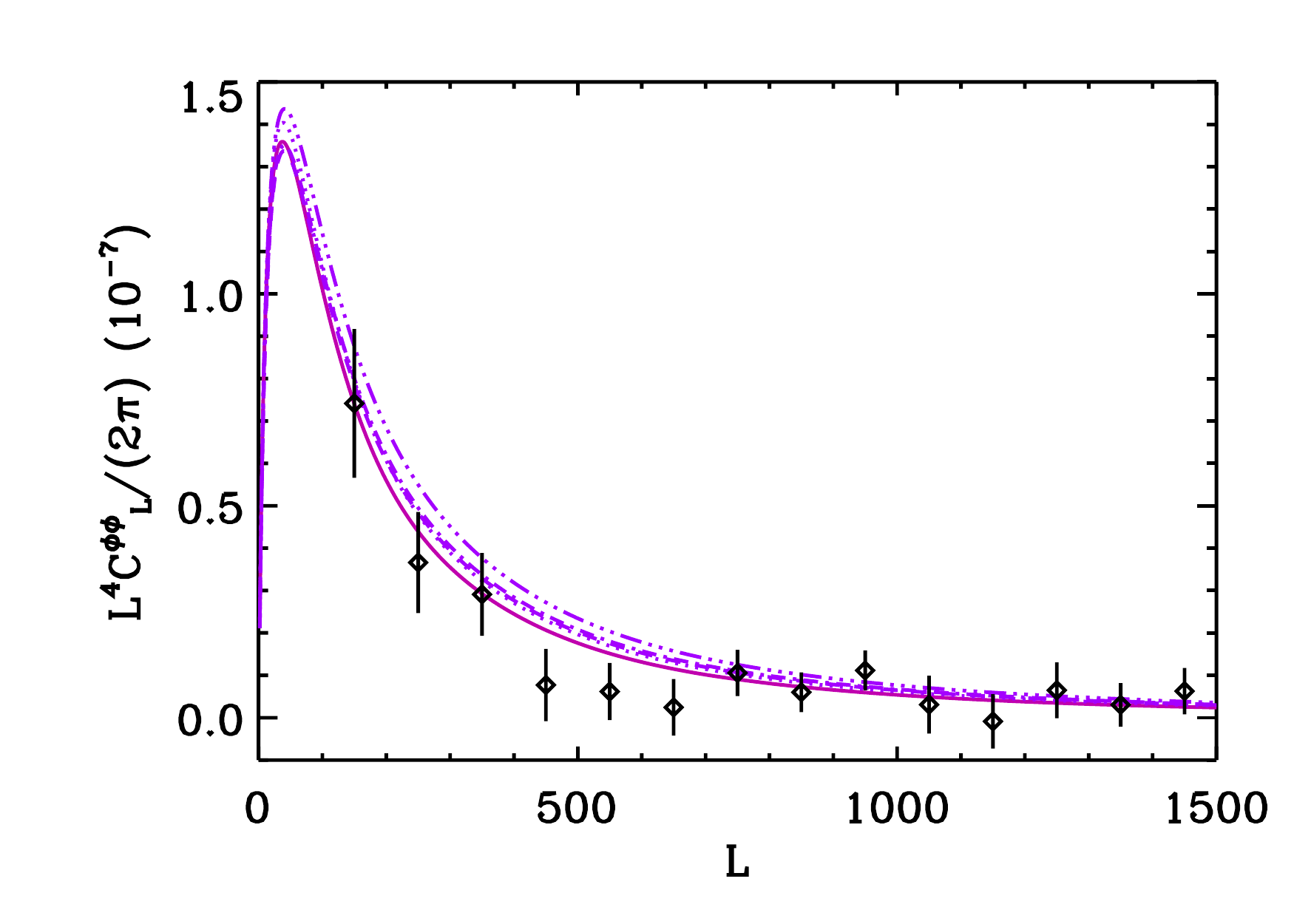}
\includegraphics[width=7.5cm]{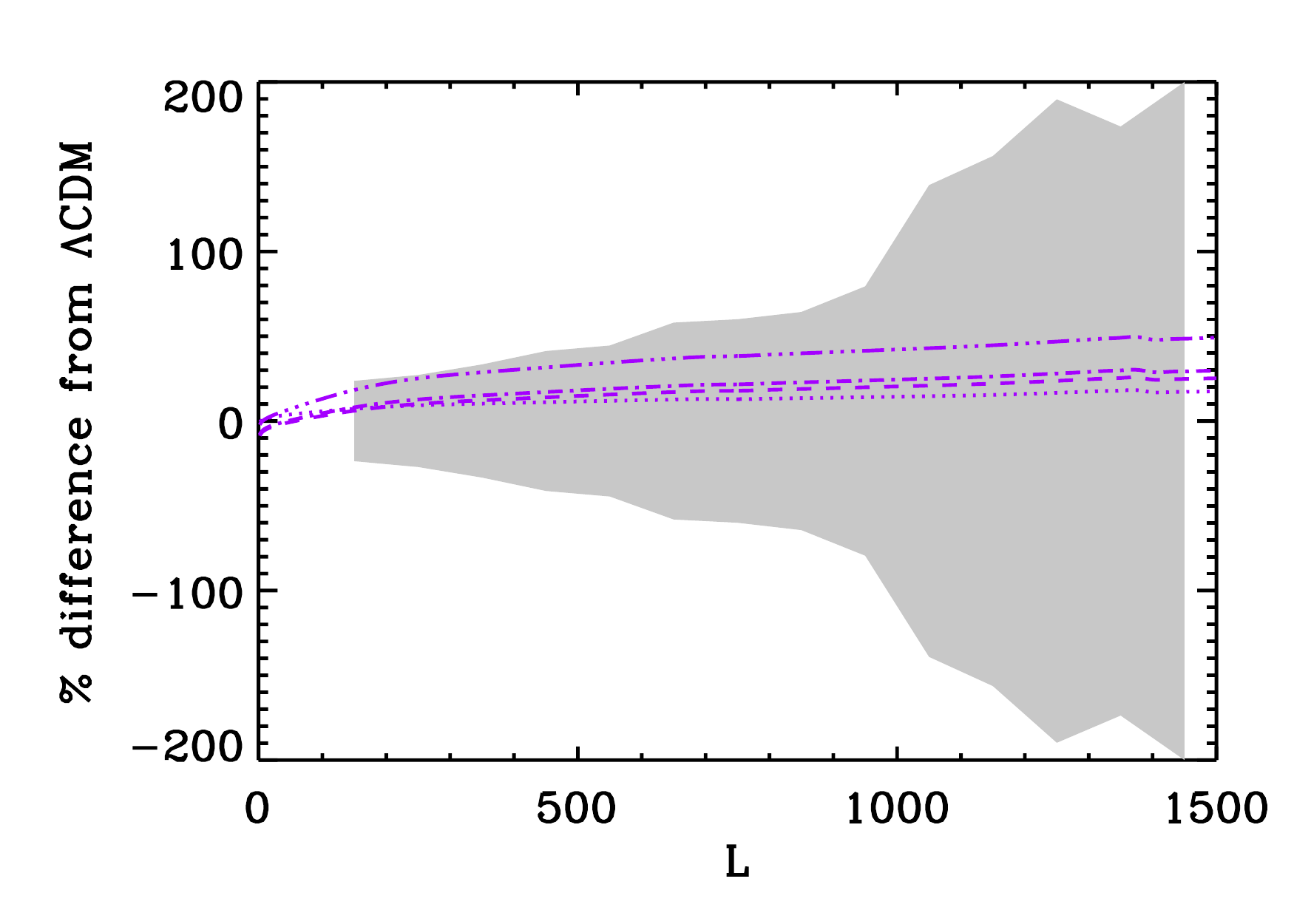}
\caption{Left: CMB lensing data from the South Pole Telescope (black diamonds). Overlaid are power spectra for $\Lambda$CDM (solid) and for nearly-degenerate models with $\neff$ equal to $4.0$ (dotted), $5.0$ (dashed), $5.8$ (dot-dashed) and $6.0$ (triple-dot-dashed). Right: the fractional differences of the power spectra with respect to $\Lambda$CDM are shown as a function of scale, with SPT error bars indicated by the grey shaded region.}
\label{fig:CMBlensCl}
\end{figure}

We consider the datasets listed below in our analysis.
\begin{itemize}
\item Seven-year data from the Wilkinson Microwave Anisotropy Probe (WMAP) satellite~\cite{Larson:2010gs}. In particular we consider the angular power spectra ($C_{\ell}$) of the temperature and E-mode  polarization signals and their cross-correlation. We use the likelihood routine released by the WMAP collaboration, which is available for download from the Legacy Archive for Microwave Background Data Analysis (LAMBDA) website.\footnote{\url{http://lambda.gsfc.nasa.gov/product/map/dr4/m_products.cfm}}
\item The South Pole Telescope (SPT) measurement of the damping tail of the CMB temperature $C_{\ell}$. We follow Ref.~\cite{Keisler:2011aw} in computing the SPT likelihood, setting $\ell_\mathrm{max} = 3000$, and marginalizing over foreground contributions from unresolved point sources and Sunyaev-Zel'dovich (SZ) clusters.
\item The South Pole Telescope measurement of the signature of weak gravitational lensing of the CMB by large-scale structure, yielding an estimate of the power spectrum of the projected gravitational potential~\cite{vanEngelen:2012va} (SPTLens). In the Appendix we describe in detail how this likelihood is computed. We make the Fortran likelihood code written for this work publicly available for download.\footnote{\url{http://zuserver2.star.ucl.ac.uk/~smf/spt_lensing_likelihood.tar.gz}}
\item Measurements of the expansion history of the Universe at low redshifts ($z<1$) using observations of the baryon acoustic oscillation scale (BAO) by WiggleZ~\cite{Blake:2011en} (at $\langle z \rangle= \{0.44,\, 0.6,\, 0.73 \}$) and the Sloan Digital Sky Survey~\cite{Percival:2009xn} (at $\langle z \rangle = \{ 0.2,\,0.35 \}$).
\item A measurement of  $H_0$ from Ref.~\cite{Riess:2009pu}.
\end{itemize}
The CMB temperature and lensing data we use are shown in the left-hand plots of Figs.~\ref{fig:CMBCl} and~\ref{fig:CMBlensCl}.\footnote{Note that the CMB lensing power spectra are plotted as a function of $L$, rather than $\ell$. As CMB lensing induces couplings between modes with different $\ell$ values, the signal is estimated as a function of the vector ${\bf L}$, the sum of two harmonic-space CMB modes: ${\bf L} = {\bf l}_1 + {\bf l}_2$. The magnitude of this quantity, $L$, is still an inverse angular scale, and lensing power spectra can thus be interpreted in exactly the same way as temperature power spectra.} Note that to make the damping tail more visible we have plotted the angular temperature power spectrum in the form of $\ell^3(\ell+1)C_{\ell}/(2\pi)$.

In our analysis we use WMAP data alone only in few reference cases, as the degeneracies between $N_{\rm eff}$ and other parameters are not resolved with WMAP data alone. We also use the combinations WMAP+SPT, WMAP+SPT+SPTLens, WMAP+SPT+SPTLens+BAO, and WMAP+SPT+ SPTLens+$H_0$. These datasets represent the state of the art circa mid 2012. Since then new cosmological data -- notably new WMAP~\cite{Bennett_etal:2012}, BAO~\cite{Anderson_etal:2012}, SPT~\cite{Story_etal:2012} and {\em Planck}~\cite{Planck_Overview:2013} results -- have become public, which we have not included for several reasons. The first, and purely practical, consideration is that each Evidence run takes a significant amount of time, and one is therefore forced to ``freeze'' the data considered early in the analysis. Furthermore:
\begin{itemize}
\item The WMAP9 data are fully in agreement with the previous release and, coupled with the fact that we use WMAP data alone in only a few selected cases, the difference between using WMAP7 and WMAP9 data is negligible.
\item The impact of BAO data on our findings is, at the moment, only marginal, as already noted in Ref.~\cite{Hou:2011ec}. We therefore argue that the increased statistical power of new BAO data, especially in combination with other datasets, will not significantly change our findings. Moreover, as the authors of the relevant papers point out (in, for example, Sec. 8 of Ref.~\cite{Anderson_etal:2012}), there seems to be a hint of tension between BAO and CMB data, which we will discuss further below.
\item The new SPT data were released late in the process of writing this manuscript, and also exhibit a mild tension with other datasets~\cite{diValentino_etal:2013}. Since the source of this tension is unknown at present, and systematics may potentially play a role, we do not include these new data in this work.
\item The {\em Planck} data were released after this work was submitted for publication. An analysis including the {\em Planck} data will be presented in forthcoming work~\cite{Feeney_etal:2013}.
\end{itemize}


\subsection{Models and Priors}
\label{sec:priors}

The models considered in this work are listed below. In all cases, the geometry is assumed to be flat.
\begin{itemize}
\item A baseline power-law $\Lambda$CDM model with three massless neutrinos ($N_{\rm eff}=3.046$): $\Lambda$CDM.
\item A power-law $\Lambda$CDM model with massless neutrinos in which $N_{\rm eff}$ is allowed to vary: $\Lambda$CDM$\neff$.
\item A power-law $\Lambda$CDM model with three neutrinos with non-zero mass:  $\Lambda$CDM$\mnu$.
\item A power-law $\Lambda$CDM model with massive neutrinos in which $N_{\rm eff}$ is allowed to vary: $\Lambda$CDM$\mnu\neff$.
\item A power-law $\Lambda$CDM model with massless neutrinos in which $N_{\rm eff}$ is allowed to vary and the power-law spectral index is allowed to run: $\nrun\Lambda$CDM$\neff$.
\end{itemize}

We use a uniform prior on the following parameters: the physical cold dark matter density, $\Omega_c h^2$; the physical baryon density, $\Omega_b h^2$; $\exp(-2\tau)$ ($\tau$ being the optical depth to the last scattering surface); the power spectrum slope, $n_s$; the running of the spectral index,  $\nrun = {\rm d}n_s / {\rm d}\ln k$, where $k$ is the wavenumber; and the log of the scalar amplitude, $\log A_s$. For these parameters the range of the prior does not affect the Evidence calculation for neutrino properties as long as the prior encompasses the region where the likelihood is significant, which is the case here. We also use a uniform prior on the parameter $\theta$ because it roughly linearizes the dependence of the $C_\ell$s on the cosmological parameters~\cite{Kosowsky_etal:2002}, and is therefore a good sampling parameter. In standard $\Lambda$CDM or $\nrun\Lambda$CDM, $\theta$ is the  the ratio of the  sound horizon to the angular diameter distance at decoupling. {\tt CosmoMC} computes this quantity  approximately, using the fitting formula of Ref.~\cite{Hu:1994uz} to compute the redshift of decoupling, which is not valid for models with, for example, non-standard neutrinos. The approximation for the redshift of decoupling is good to 4\% in models with non-standard neutrinos. Thus in such models the parameter $\theta$ is only approximately the ratio of the sound horizon to the angular diameter distance.  The actual calculation for CMB anisotropies is done correctly using the parameter $\hnought$: by sampling uniformly in the parameter $\theta$ computed with the approximation from Ref.~\cite{Hu:1994uz} one is effectively using a slightly non-uniform prior on the true ratio of the sound horizon to the angular diameter distance. Some authors have avoided this problem by sampling uniformly in $H_0$. However, $H_0$ and $\theta$ are heavily degenerate with, for example, $\neff$ and so the priors imposed on these parameters directly affect the calculation of the Evidence.

We assume a uniform prior on $N_{\rm eff}$ in the range $1.047 \le \neff \le 9.0$. The upper limit is chosen to match that used in the SPT analysis of Ref.~\cite{vanEngelen:2012va}. The lower limit is imposed by the fact that the approximation used by {\tt CosmoMC} to calculate the redshift of decoupling~\cite{Hu:1994uz} breaks down at this point. In the majority of cases, the likelihood is effectively zero near the minimum value of $\neff$ considered, and no contribution to the Evidence is therefore missed. Truncating the $\neff$ prior at $1.047$ instead of zero increases the log-Evidence (via the reduction in prior volume) by $0.12$: as this is smaller than the typical sampling precision in calculating the log-Evidence ($0.3$) we ignore this effect.

The variation of $\mnu$ is achieved by sampling the fraction, $\fnu$, of the dark matter that is in neutrinos~\cite{Spergel_etal:2003}
\begin{equation}
\fnu = \frac{ \omgnuh }{ \omgmh } = \frac{ 1 }{ \omgmh } \frac{ \mnu }{ 94 {\rm \,eV} }.\footnote{The factor of 94 eV relates the energy density and total mass of $N_\nu$ neutrino species with a common temperature. When simultaneously sampling $\neff$ and $\fnu$, $\mnu$ is only strictly the sum of the neutrino masses if the extra neutrinos are assumed to be at the same temperature as the Standard Model neutrinos.}
\end{equation}
We assume a uniform prior on $\fnu$ in the range $0 \le \fnu \le 0.43$. The lower limit corresponds to massless neutrinos; the upper limit is the maximum contribution neutrinos would make to the dark matter assuming $\mnu \sim 2$ eV and the minimum physical dark matter density allowed by the prior. The priors we adopt on $\fnu$ and $\omgmh$ allow the neutrino mass to range between $0.0 \le \mnu \le 8$ eV, slightly larger than the upper limit given by anti-electron-neutrino mass measurements~\cite{Troitsk:2011}.

In Table~\ref{tab:runs} we present the combinations of datasets and models we consider. 

\begin{table}[tb]
\centering
 \begin{tabular}{||l ||c| c| c| c| c|| }
  \hline
  \hline
  models/data & WMAP & WMAP+SPT& WMAP+SPT &   WMAP+SPT & WMAP+SPT\\
  & & & +SPTLens & +SPTLens+BAO & +SPTLens+$H_0$\\
  \hline
  $\Lambda$CDM & E & E & E & E & E \\
  \hline
  $\Lambda$CDM$\neff$ & E & E & E & E & E \\
 \hline
 $\Lambda$CDM$\mnu$ & E & E & E & E & E \\
 \hline
 $\Lambda$CDM$\mnu\neff$ & & E & E & E & E\\
 \hline
 \hline
 $\nrun\Lambda$CDM$\neff$ & & E & E &  &  \\
\hline
\hline
\end{tabular}
 \caption{Summary of the combinations of models and datasets for which the Evidence is calculated (E). For the abbreviations used, see the main text.}
 \label{tab:runs}
\end{table}


\section{Results}
\label{sec:results}

We begin by presenting the parameter constraints, then move to model selection and report the Evidence for deviations from Standard Model neutrinos and the profile likelihood ratio.


\subsection{Parameter Estimation}
\label{sec:res_params}

Parameter estimates for the various models and data combinations considered are shown in Table~\ref{tab:constraints}. Constraints on the summed neutrino mass, $\mnu$, are presented as $95\%$ confidence upper limits; all other parameter estimates are mean-posterior values with $68\%$ confidence-limit errors. Concentrating first on the $\Lambda$CDM$\neff$ model, we see that, for the majority of data combinations, the mean posterior value of $\neff$ is higher than the Standard Model prediction by approximately $1.5\sigma$. We note that the SPT lensing data do not contribute significantly to the constraints. While in principle the CMB lensing signal is sensitive to cosmological parameters such as $\neff$ and $\mnu$, the use of the signal is in its infancy, having been detected for the first time only in the last year or so. The error bars are still too large to contribute significantly to parameter constraints, but a reduction of the error-bar magnitude by a factor of few, especially on large-scales ($L < 500$), should change this (as can be appreciated from the right-hand panel of Fig.~\ref{fig:CMBlensCl}). Increased sky coverage could easily reduce the error bars by the required factor.

 \begin{table}[tb]
\centering
\tabcolsep=0.11cm
{\small
 \begin{tabular}{|| c | c | c | c | c | c | c | c ||}
  \hline
  \hline
  model and data & $\omgmh$ & $\hnought$ & $100 \times \yhe$ & $\sigeight$ & $\neff$ & $\mnu$ & $\nrun$ \\
  \hline
   & & & & & & &  \\
  $\Lambda$CDM & & & & & & &  \\
  \hline
  WMAP & $0.113 \pm 0.006$ & $70 \pm 2$ & $24.78 \pm 0.02$ & $0.82 \pm 0.03$ & - & - & - \\
  WMAP+SPT & $0.112 \pm 0.005$ & $71 \pm 2$ & $24.78 \pm 0.02$ & $0.81 \pm 0.02$ & - & - & - \\
  WMAP+SPT+SPTLens & $0.112 \pm 0.005 $ & $71 \pm 2$ & $24.78 \pm 0.02$ & $0.81 \pm 0.02$ & - & - & - \\
  WMAP+SPT+SPTLens+BAO & $0.115 \pm 0.003$ & $69 \pm 1$ & $24.78 \pm 0.02$ & $0.82 \pm 0.02$ & - & - & - \\
  WMAP+SPT+SPTLens+$\hnought$ & $0.109 \pm 0.004$ & $72 \pm 2$ & $24.78 \pm 0.02$ & $0.80 \pm 0.02$ & - & - & - \\
  \hline
   & & & & & & &  \\
  $\Lambda$CDM$\neff$ & & & & & & &  \\
  \hline
  WMAP & $0.15 \pm 0.03$ & $79 \pm 7$ & $27 \pm 2$ & $0.90 \pm 0.07$ & $5 \pm 2 $ & - & - \\
  WMAP+SPT & $0.13 \pm 0.01$ & $75 \pm 4$ & $25.8 \pm 0.7$ & $0.86 \pm 0.04$ & $3.9 \pm 0.6$ & - & - \\
  WMAP+SPT+SPTLens & $0.13 \pm 0.01$ & $75 \pm 4$ & $25.8 \pm 0.7$ & $0.86 \pm 0.04$ & $3.9 \pm 0.6$ & - & - \\
  WMAP+SPT+SPTLens+BAO & $0.13 \pm 0.01$ & $73 \pm 3$ & $25.8 \pm 0.7$ & $0.87 \pm 0.04$ & $3.8 \pm 0.6$ & - & - \\
  WMAP+SPT+SPTLens+$\hnought$ & $0.12 \pm 0.01$ & $74 \pm 2$ & $25.7 \pm 0.5$ & $0.85 \pm 0.04$ & $3.7 \pm 0.4$ & - & - \\
  \hline
   & & & & & & &  \\
  $\Lambda$CDM$\mnu$ & & & & & & &  \\
  \hline
  WMAP & $0.119 \pm 0.007$ & $65 \pm 4$ & $24.77 \pm 0.03$ & $0.72 \pm 0.07$ & - & $< 1.1$ & - \\
  WMAP+SPT & $0.120 \pm 0.007$ & $64 \pm 4$ & $24.76 \pm 0.02$ & $0.69 \pm 0.07$ & - & $< 1.3$ & - \\
  WMAP+SPT+SPTLens & $0.121 \pm 0.008$ & $64 \pm 4$ & $24.76 \pm 0.02$ & $0.69 \pm 0.07$ & - & $< 1.3$ & - \\
  WMAP+SPT+SPTLens+BAO & $0.115 \pm 0.003$ & $68 \pm 2$ & $24.77 \pm 0.02$ & $0.74 \pm 0.05$ & - & $< 0.7$ & - \\
  WMAP+SPT+SPTLens+$\hnought$ & $0.110 \pm 0.004$ & $71 \pm 2$ & $24.78 \pm 0.02$ & $0.76 \pm 0.04$ & - & $< 0.4$ & - \\
  \hline
   & & & & & & &  \\
  $\Lambda$CDM$\mnu\neff$ & & & & & & &  \\
  \hline
  WMAP+SPT & $0.13 \pm 0.01$ & $68 \pm 6$ & $25.6 \pm 0.8$ & $0.74 \pm 0.08$ & $3.7 \pm 0.6$ & $< 1.3$ & - \\
  WMAP+SPT+SPTLens & $0.13 \pm 0.01$ & $68 \pm 6$ & $25.6 \pm 0.8$ & $0.74 \pm 0.08$ & $3.7 \pm 0.6$ & $< 1.3$ & - \\
  WMAP+SPT+SPTLens+BAO & $0.13 \pm 0.01$ & $71 \pm 3$ & $25.7 \pm 0.7$ & $0.78 \pm 0.07$ & $3.8 \pm 0.6$ & $< 0.7$ & - \\
  WMAP+SPT+SPTLens+$\hnought$ & $0.13 \pm 0.01$ & $73 \pm 2$ & $25.9\pm  0.6$ & $0.80 \pm 0.05$ & $4.0 \pm 0.5$ & $< 0.7$ & - \\
  \hline
   & & & & & & &  \\
  $\nrun\Lambda$CDM$\neff$ & & & & & & &  \\
  \hline
  WMAP+SPT & $0.12 \pm 0.01$ & $69 \pm 6$ & $25 \pm 1$ & $0.83 \pm 0.05$ & $3.0 \pm 0.9$ & - & $-0.02 \pm 0.02$ \\
  WMAP+SPT+SPTLens & $0.12 \pm 0.01$ & $69 \pm 6$ & $25 \pm 1$ & $0.83 \pm 0.04$ & $3.0 \pm 0.8$ & - & $-0.02 \pm 0.02$ \\
  \hline
  \hline
\end{tabular}
}
 \caption{Parameter-estimation results for the various model-dataset combinations. All parameter estimates are presented as mean-posterior values with $68\%$ CL errors, apart from the constraints on the summed neutrino mass, $\mnu$, which are $95\%$ confidence upper limits. Almost all dataset combinations favor $\neff > 3.046$ at the $\sim 1.5\sigma$ level, unless the primordial power spectrum is allowed to run ($\alpha\Lambda$CDM). Degeneracies between $\neff$ and $\omgmh$, $H_0$, $\yhe$ and $\nrun$, and between $\mnu$ and $\sigeight$, are clear from the increased uncertainty in the standard cosmological parameters when the relevant neutrino parameter is allowed to vary.
 \label{tab:constraints}}
\end{table}

Increasing the amount of relativistic matter in the early Universe affects the expansion history, and in particular, the redshift of matter-radiation equality, $\zeq$. The effects of additional massless neutrinos -- or equivalently dark radiation -- are therefore degenerate with parameters such as the physical dark matter density, $\omgmh$. CMB temperature power spectra for two cosmologies with the same $1+\zeq$ as $\Lambda$CDM but non-standard $\neff$ are plotted in Fig.~\ref{fig:CMBCl} alongside the WMAP and SPT measurements and errors. The raw power spectra (left) are nearly indistinguishable by eye; the differences are only apparent from the fractional difference plot (right), in which the measurement errors are plotted as grey shaded regions: the power spectra are all compatible with CMB experiments to-date. Equivalent plots for the CMB lensing power spectra are shown in Fig.~\ref{fig:CMBlensCl}.

\begin{figure}[tb]
\centering
\includegraphics[width=7.5cm]{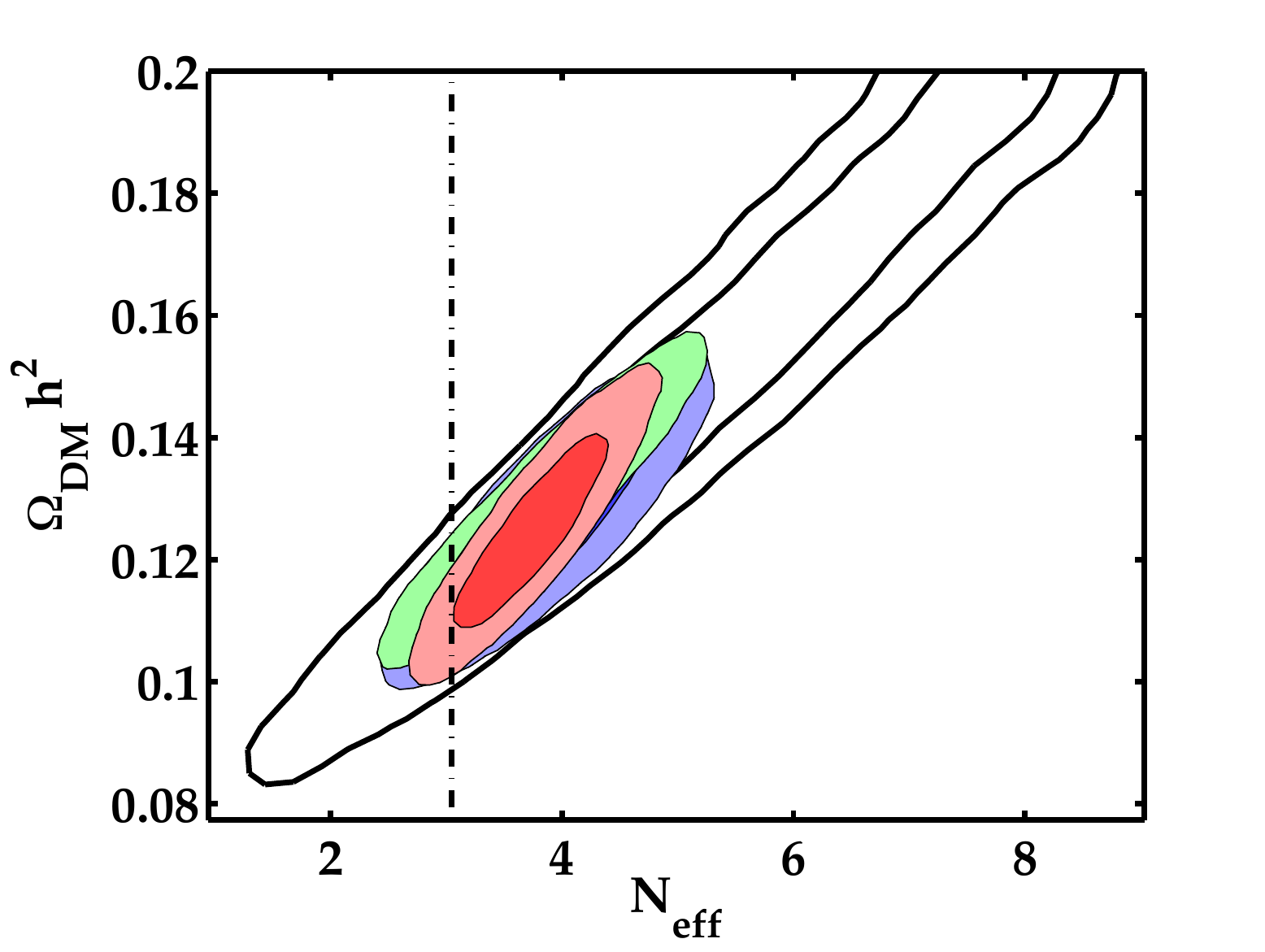}
\includegraphics[width=7.5cm]{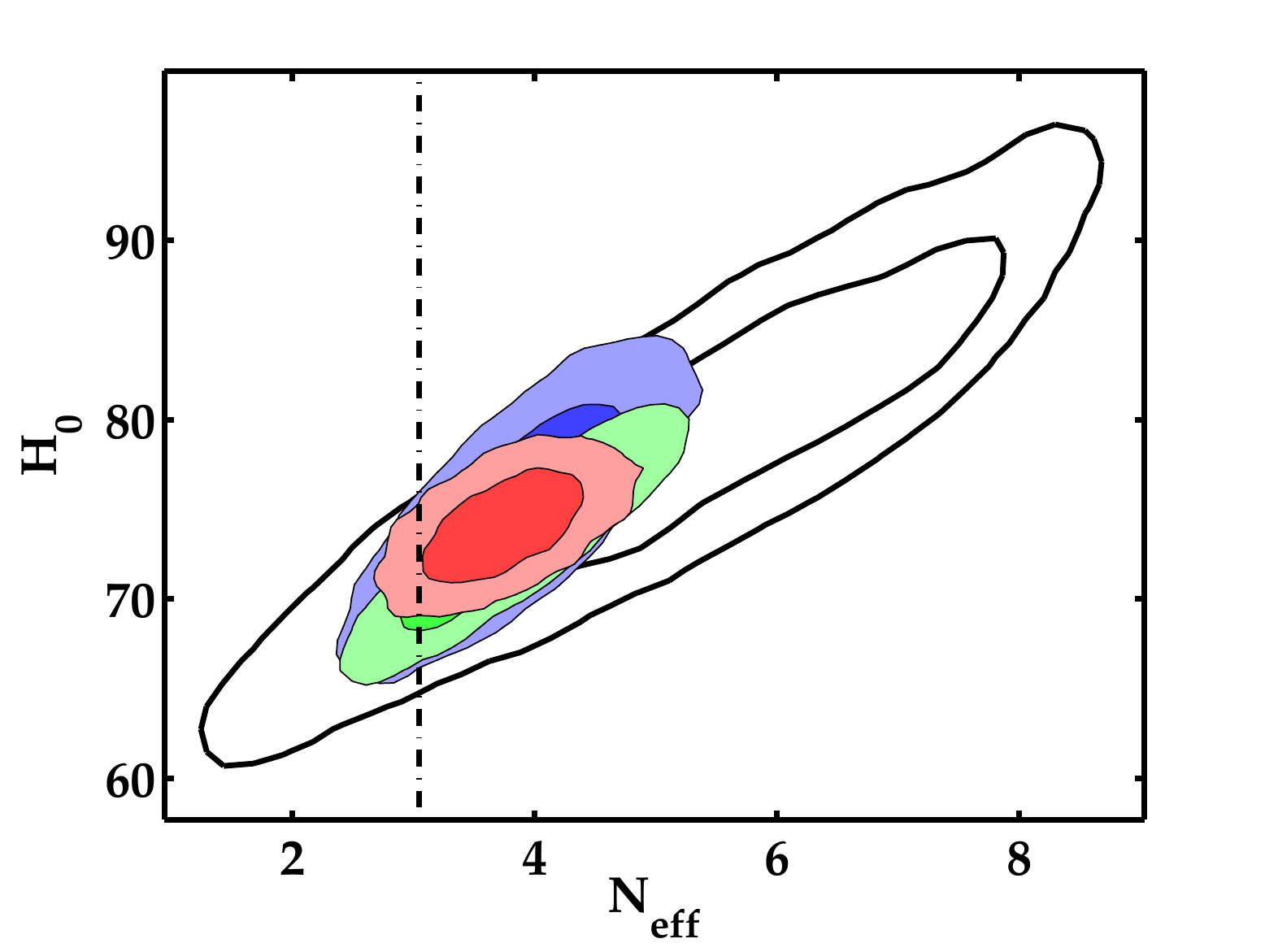}
\caption{The $68\%$ and $95\%$ joint confidence limits on the posterior probability distributions of $\neff$ and $\omgmh$ (left), and $\neff$ and $\hnought$ (right), given the WMAP (clear), WMAP+SPT+SPTLens (blue), WMAP+SPT+SPTLens+BAO (green) and WMAP+SPT+SPTLens+$\hnought$ (red) datasets. Constraints from the WMAP+SPT datasets are not shown as they are very similar to results from the WMAP+SPT+SPTLens combination. The standard value of $\neff$ is indicated by the dot-dashed line.}
\label{fig:neff_neff_vs_omegadmh2_h0}
\end{figure}

\begin{figure}[tb]
\centering
\includegraphics[width=7.5cm]{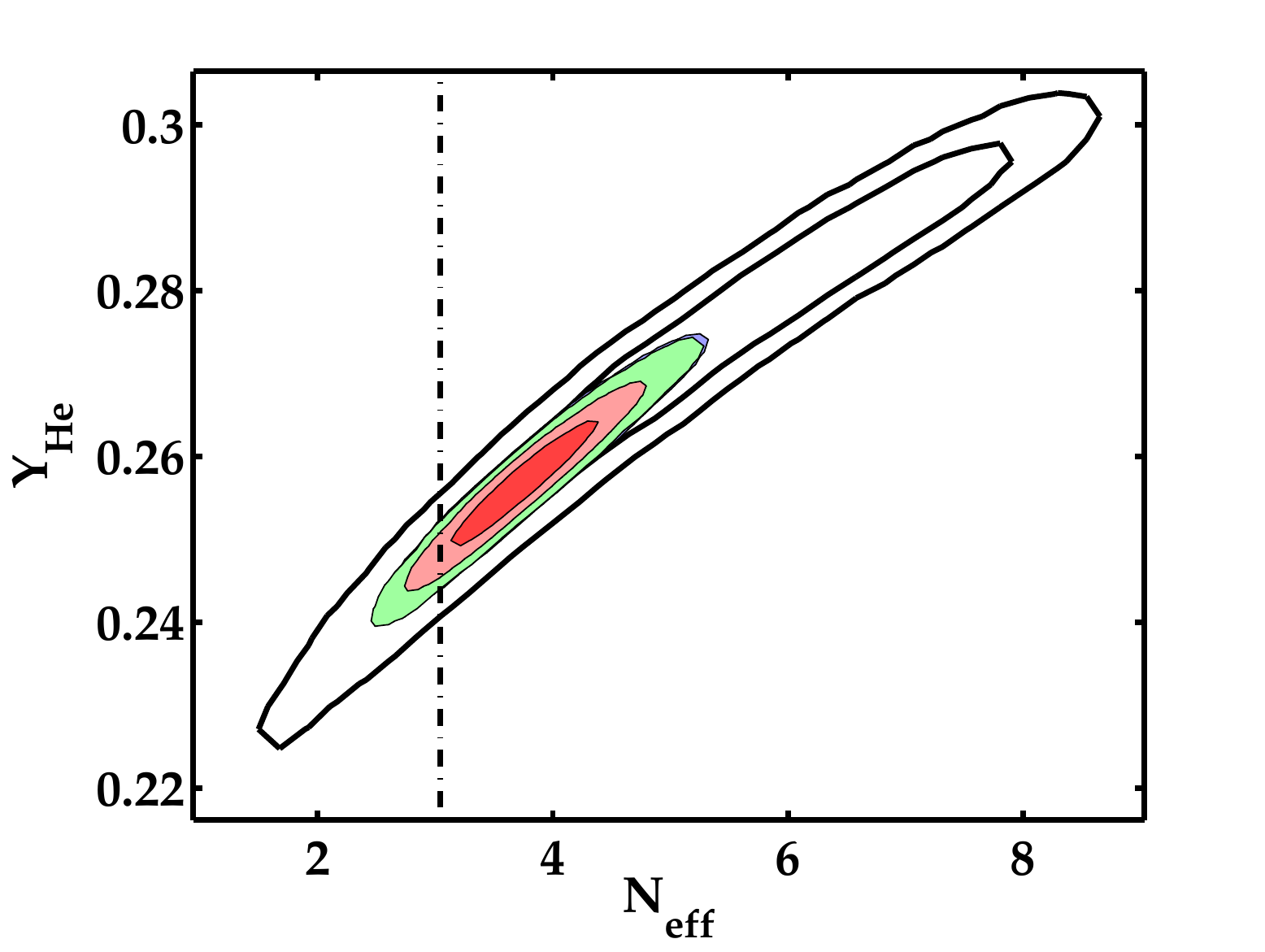}
\includegraphics[width=7.5cm]{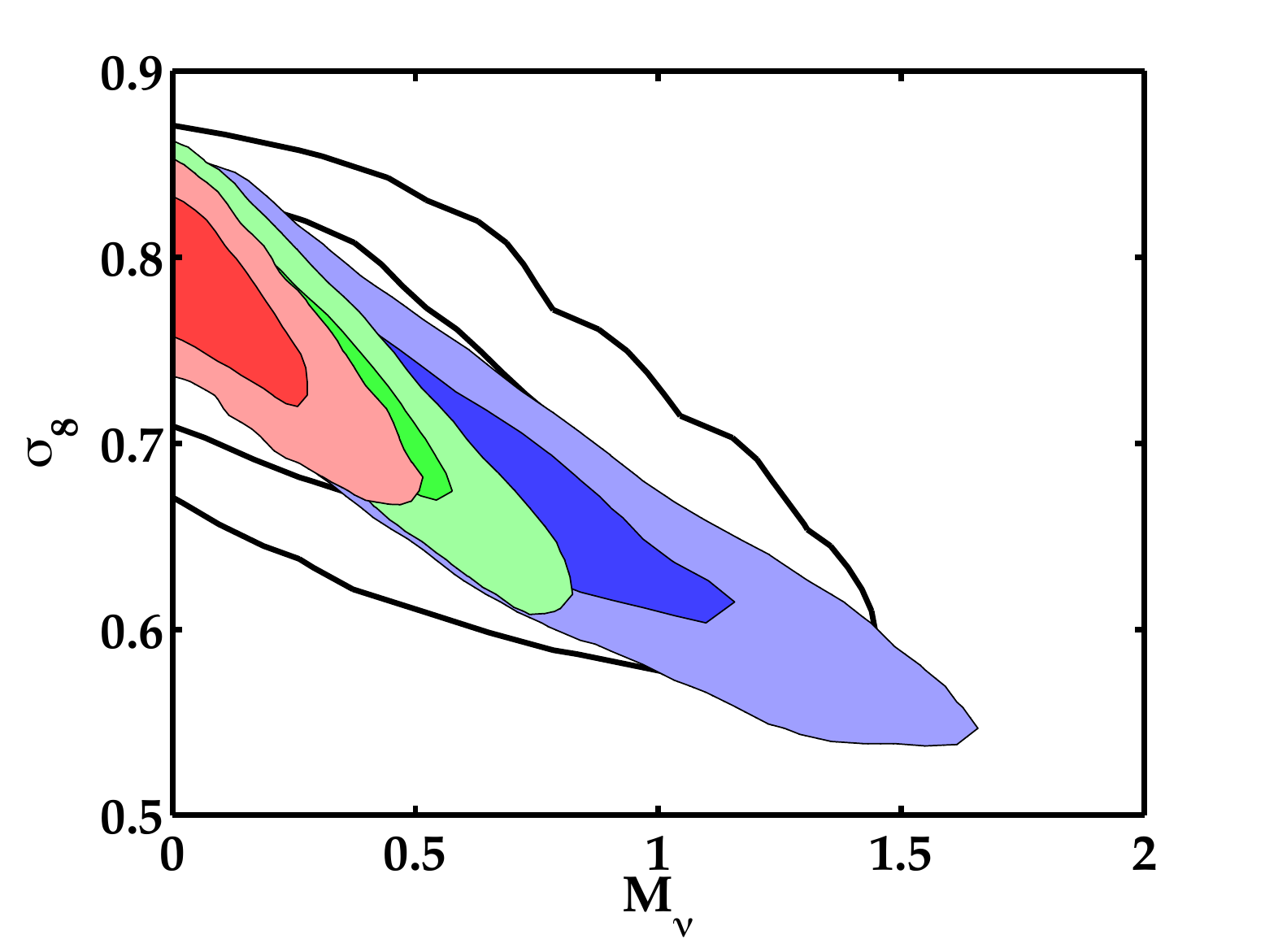}
\caption{The $68\%$ and $95\%$ joint confidence limits on the posterior probability distributions of $\neff$ and $\yhe$, and $\mnu$ (plotted in units of eV) and $\sigeight$, given the WMAP (clear), WMAP+SPT+SPTLens (blue), WMAP+SPT+SPTLens+BAO (green) and WMAP+SPT+SPTLens+$\hnought$ (red) datasets. The standard value of $\neff$ is indicated by the dot-dashed line.}
\label{fig:neff_neff_vs_yhe_fnu_mnu_vs_sigma8}
\end{figure}

The degeneracy between $\neff$ and $\omgmh$ is also clear to see in the parameter constraints plotted in Fig.~\ref{fig:neff_neff_vs_omegadmh2_h0} for all dataset combinations considered. Marginalizing over this degeneracy, which extends to large values of $\neff$, will naturally bias the posterior-mean of $\neff$ to larger values. Adding the extra data considered constrains the cosmology better, narrowing and reducing the length of the degeneracy but not breaking it. The marginalized constraint on $\neff$ is therefore not enough to claim evidence for extensions to the Standard Model. We instead turn to the Bayesian Evidence and profile likelihood ratio (in Sections~\ref{sec:res_bayes} and~\ref{sec:res_plr}, respectively) to determine whether there is any convincing evidence for a non-standard number of neutrino species. It should (and, of course, has) been noted that $\omgmh$ is not the only parameter with which $\neff$ is degenerate: the helium fraction, $\yhe$, is another such parameter~\cite{Hou:2011ec}, and $H_0$ is another. The degeneracy between these parameters is plotted in Figs.~\ref{fig:neff_neff_vs_omegadmh2_h0} and \ref{fig:neff_neff_vs_yhe_fnu_mnu_vs_sigma8}, and is clearly indicated by the estimated errors on $\yhe$ and $H_0$ increasing when $\neff$ is allowed to vary (see Table~\ref{tab:constraints}).

Parameter constraints for the models featuring a standard and non-standard number of {\em massive} neutrinos are also presented in Table~\ref{tab:constraints}. In all cases $\mnu=0$ is allowed within one-$\sigma$, and the results are therefore presented as two-$\sigma$ upper limits. Neutrinos with masses of the magnitude allowed by CMB data (i.e. $\mnu < 1$ eV) are relativistic when the CMB decouples, becoming non-relativistic at later times. While still relativistic, massive neutrinos free-stream out of, and hence damp, small-scale perturbations~\cite{Bond_etal:1980,Bond_and_Szalay:1983}. With imperfect measurements, we therefore expect $\mnu$ to be degenerate with $\sigeight$, the {\em present} linear-scale mass dispersion on a scale of $8 h^{-1}$ Mpc. As the CMB measures the power spectrum before the massive neutrinos have imparted their full suppression, $\sigeight$ and $\mnu$ are anti-correlated: the greater the contribution in massive neutrinos, the lower the present power spectrum at small scales predicted by the CMB. This degeneracy is clearly visible in Fig.~\ref{fig:neff_neff_vs_yhe_fnu_mnu_vs_sigma8}, and could be broken with an independent measurement of the small-scale power spectrum.

\begin{figure}[tb]
\centering
\includegraphics[width=7.5cm]{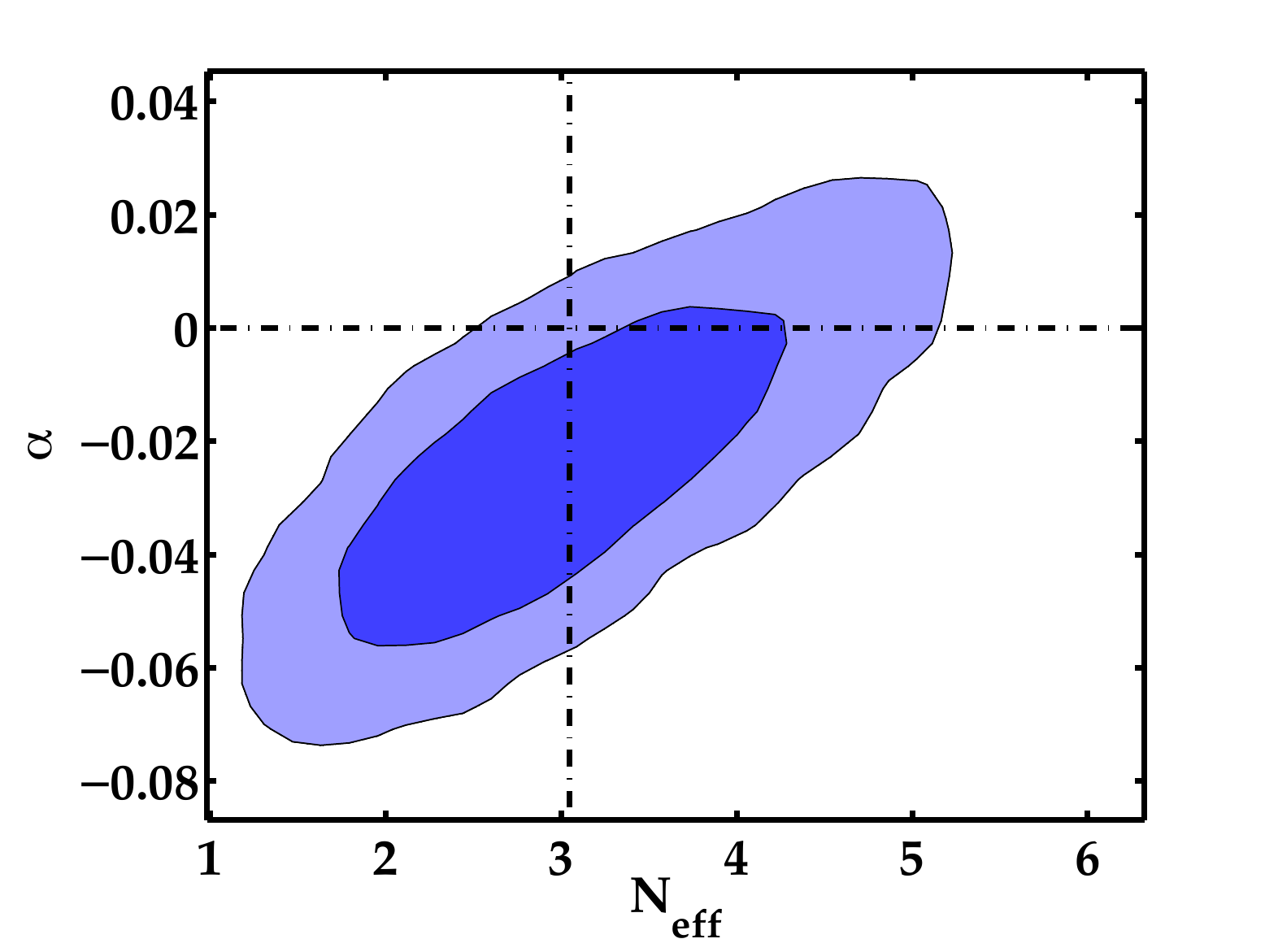}
\caption{The $68\%$ and $95\%$ joint confidence limits on the posterior probability distributions of $\neff$ and $\nrun$, given the WMAP+SPT+SPTLens (blue) datasets. The standard values of $\neff$ and $\alpha$ are indicated by the dot-dashed lines.}
\label{fig:neff_run_neff_vs_run}
\end{figure}

The final model we consider is a model with a non-standard number of massless neutrino species in which the spectral index of the scalar perturbations is allowed to run. As running can mimic the effects of additional neutrinos by suppressing small-scale power, these parameters are degenerate, as indicated by the individual (see Table~\ref{tab:constraints}) and joint (see Fig.~\ref{fig:neff_run_neff_vs_run}) parameter constraints. Nevertheless, the standard values of $\neff$ and $\nrun$ are allowed at the one-sigma level, and so there is no evidence from parameter constraints to support this model.

\subsection{Model Selection}
\label{sec:res_bayes}

\begin{figure}[tb]
\centering
\includegraphics[width=15cm]{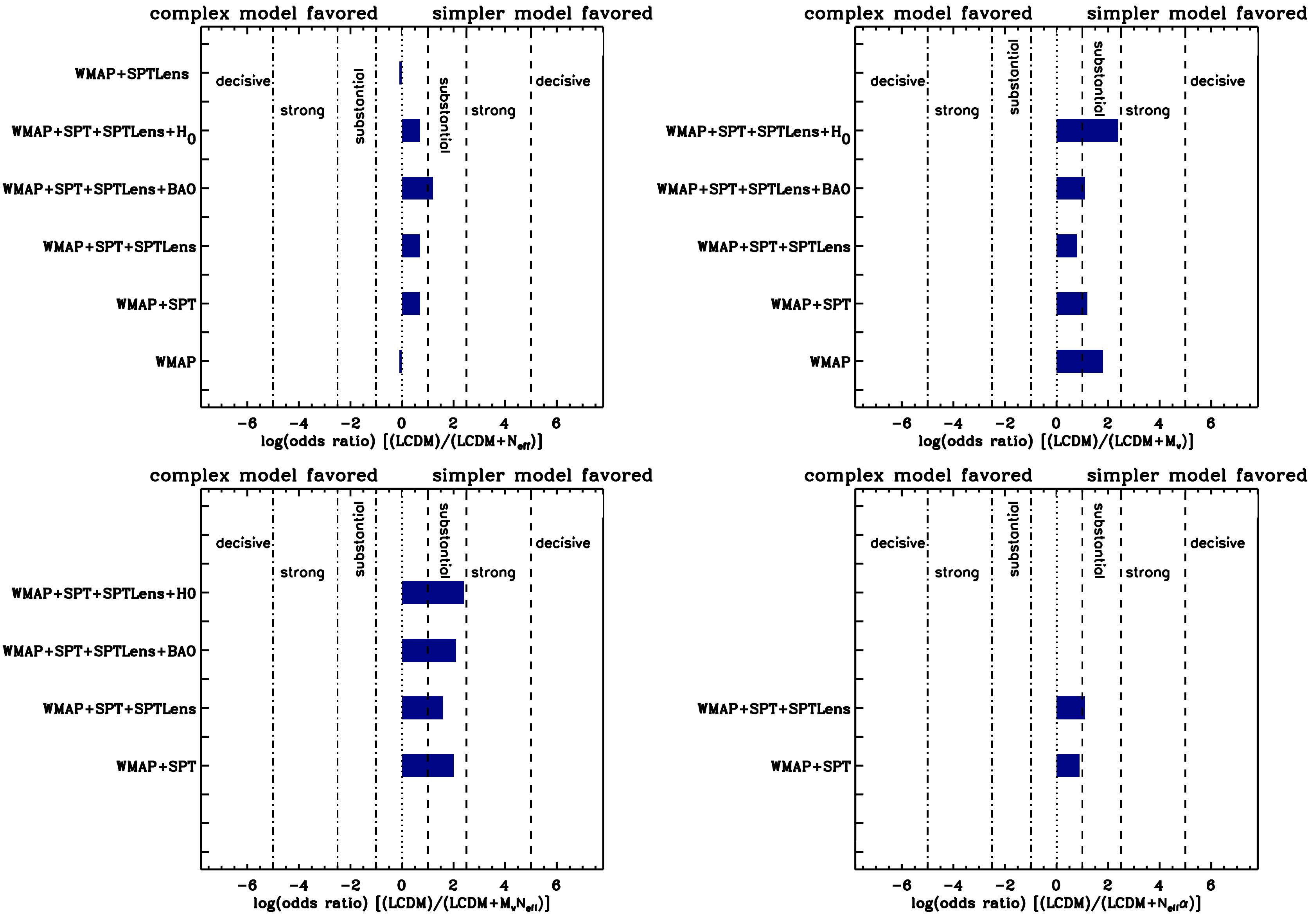}
\caption{Evidence ratios for the various extended models compared to the simple $\Lambda$CDM power-law model, for each dataset combination.
Top left: evidence for $\Lambda$CDM with a non-standard effective number of neutrino species (i.e. $\neff \ne 3.046$). Top right: evidence for adding non-zero total neutrino mass ($\mnu$) to $\Lambda$CDM. Bottom left: evidence for adding both a non-zero neutrino mass and a non-standard effective number of neutrino species to $\Lambda$CDM. Bottom right: evidence for adding a non-standard effective number of neutrino species to $\Lambda$CDM and allowing a running ($\nrun$) of the scalar spectral index. For all models, and all dataset combinations, the Evidence ratio is either inconclusive or substantially favors the simpler model, $\Lambda$CDM. The data do not favor the addition of the extra parameters considered in this analysis.}
\label{fig:evidence_plots}
\end{figure}

The results of our Evidence calculations for each model-dataset combination are presented in Fig.~\ref{fig:evidence_plots}. For each data combination, we use the Evidence for the basic $\Lambda$CDM model as the baseline for comparing models: the values plotted in Fig.~\ref{fig:evidence_plots} are
\begin{equation}
\Delta \ln E = \ln \left[ \frac{ \prob(\data | \Lambda{\rm CDM}) }{ \prob(\data | M) } \right],
\end{equation}
where $M$ is the particular extension to $\Lambda$CDM we are considering. A negative log-Evidence ratio would therefore indicate support for the more-complex model.

The immediate observation from Fig.~\ref{fig:evidence_plots} is that the Evidence is, in the majority of cases, substantially in favor of vanilla $\Lambda$CDM. This suggests that there is not yet any evidence from cosmological data for additional neutrino species or massive neutrinos. This is particularly interesting in the case of the $\Lambda$CDM$\mnu$ model, which has arguably the strongest physical motivation: neutrino oscillation measurements {\em require} neutrinos to have mass. Though the prior used in this case is somewhat {\em ad hoc} -- the limits are well-motivated, but its uniform distribution is a choice -- even by ``cheating'' (in a Bayesian sense at least, by using the data twice) and setting the maximum $\mnu$ value to that allowed by previous WMAP analyses the $\Delta \ln E$ values would only be reduced by $\sim \ln 4 \simeq1.4$. This is sufficient in some cases to tip the balance in favor of $\Lambda$CDM$\mnu$, but not with any significance.

\subsection{Profile Likelihood Ratio}
\label{sec:res_plr}

The priors applied to the additional neutrino parameters $\neff$ and $\mnu$ are, to a greater or lesser extent, phenomenologically-motivated, insofar as they are not derived from fundamental physical considerations. As we cannot therefore be confident that these priors are entirely appropriate, we employ the profile likelihood ratio (PLR) as a prior-independent, if not entirely self-consistent, model selection criterion. We concentrate on the $\Lambda$CDM$\neff$ model, as this is the only model for which the parameter estimates and Evidence ratios are in apparent disagreement.

\begin{figure}[tb]
\centering
\includegraphics[width=7.5cm]{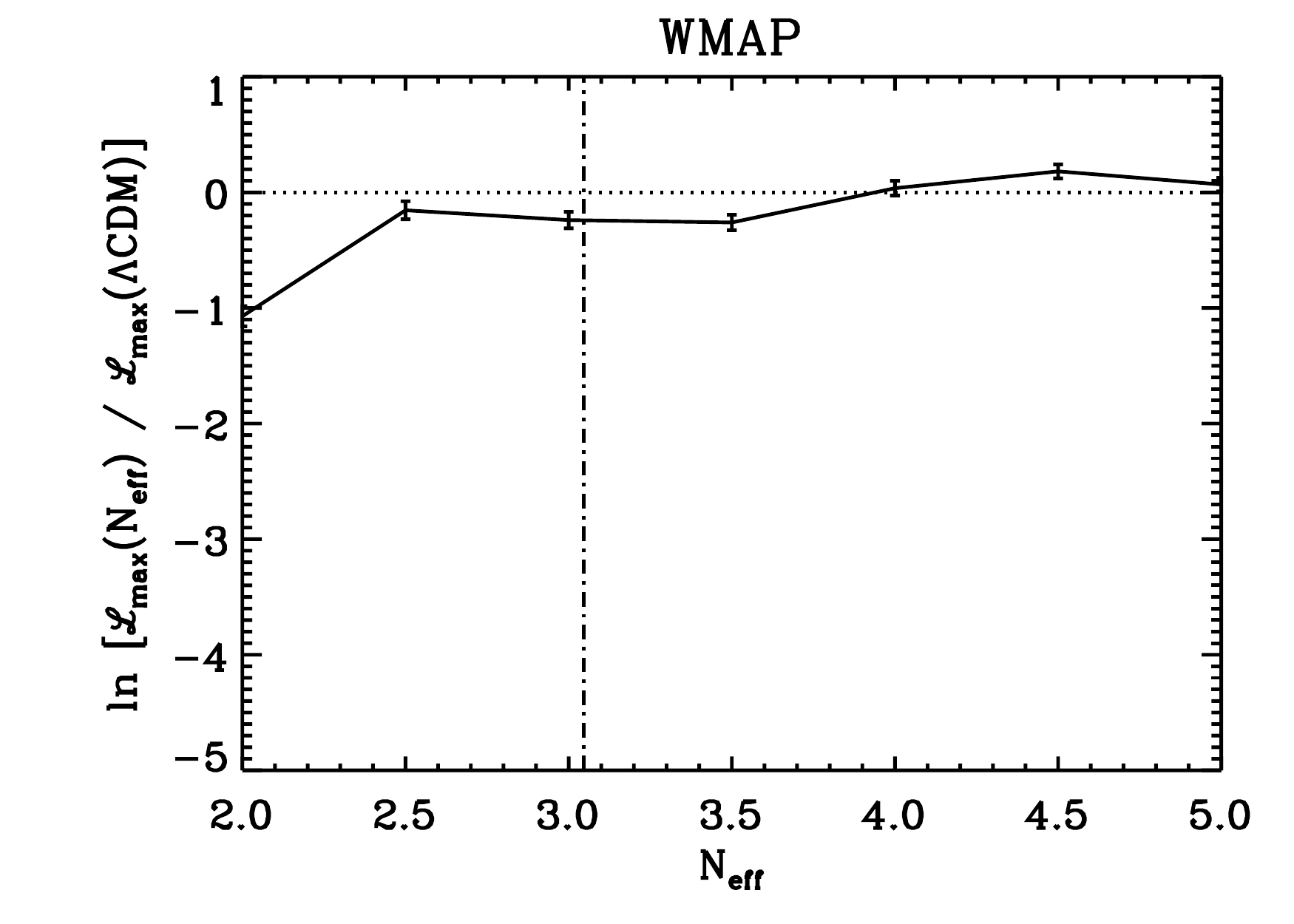}
\includegraphics[width=7.5cm]{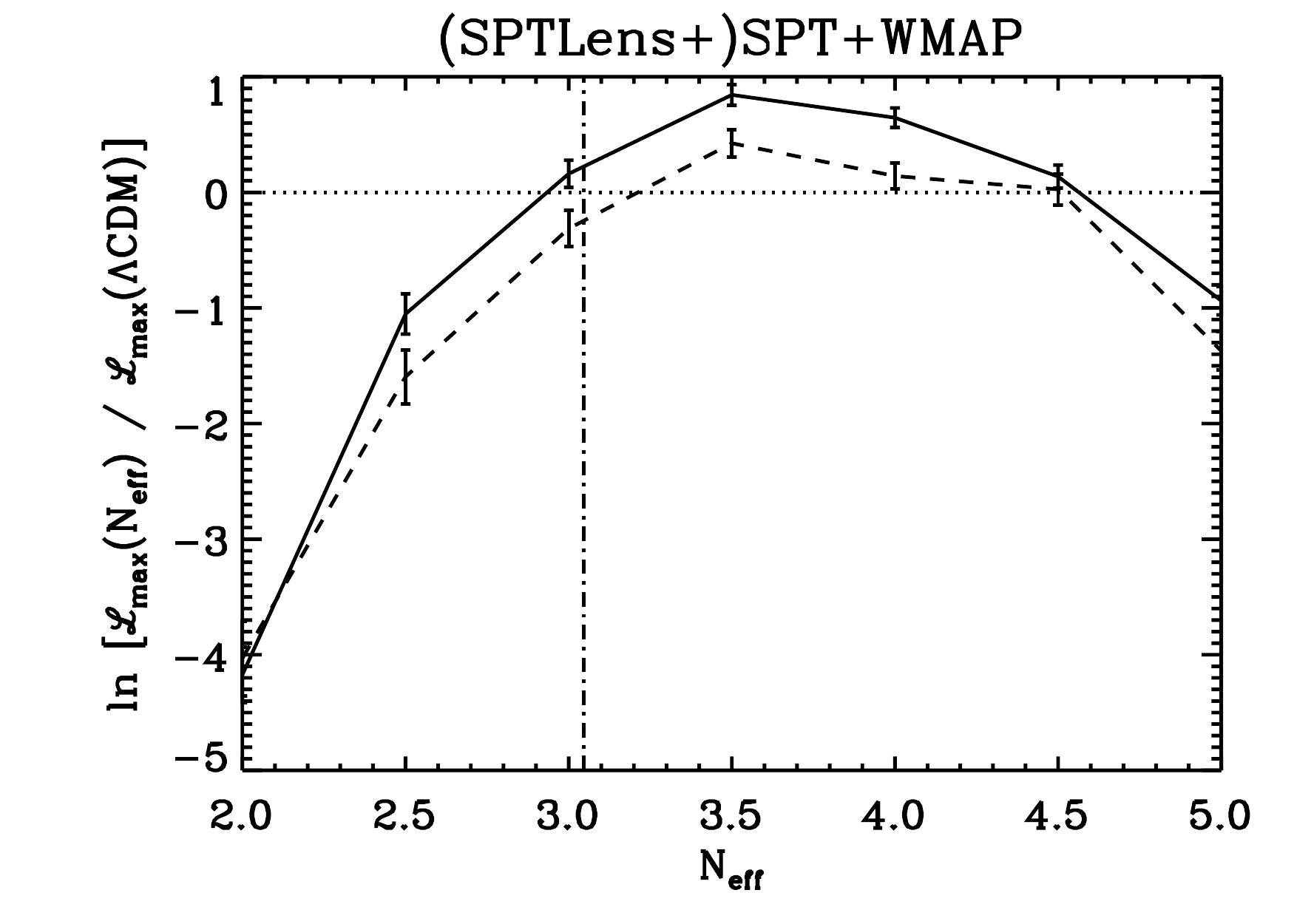}
\includegraphics[width=7.5cm]{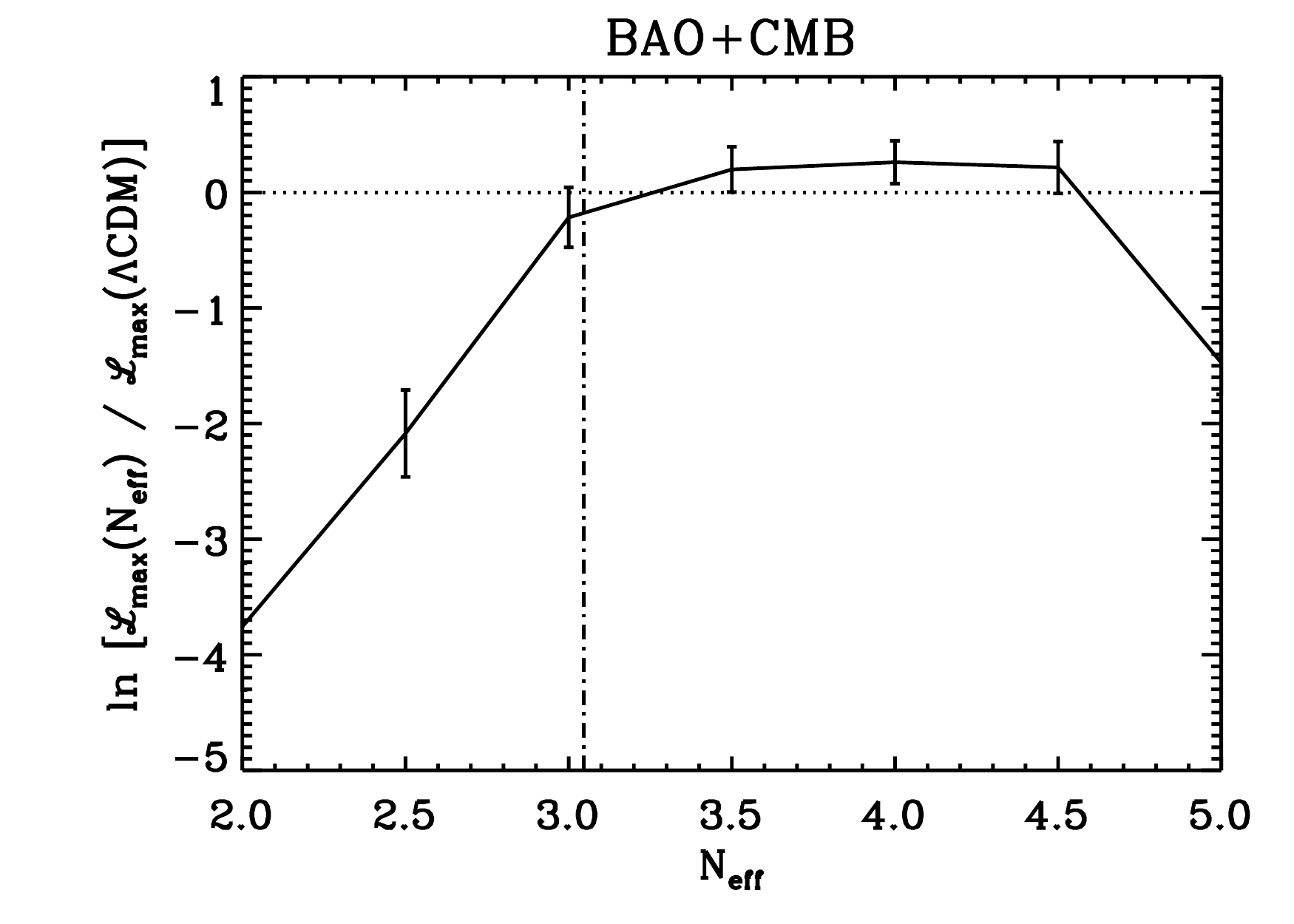}
\includegraphics[width=7.5cm]{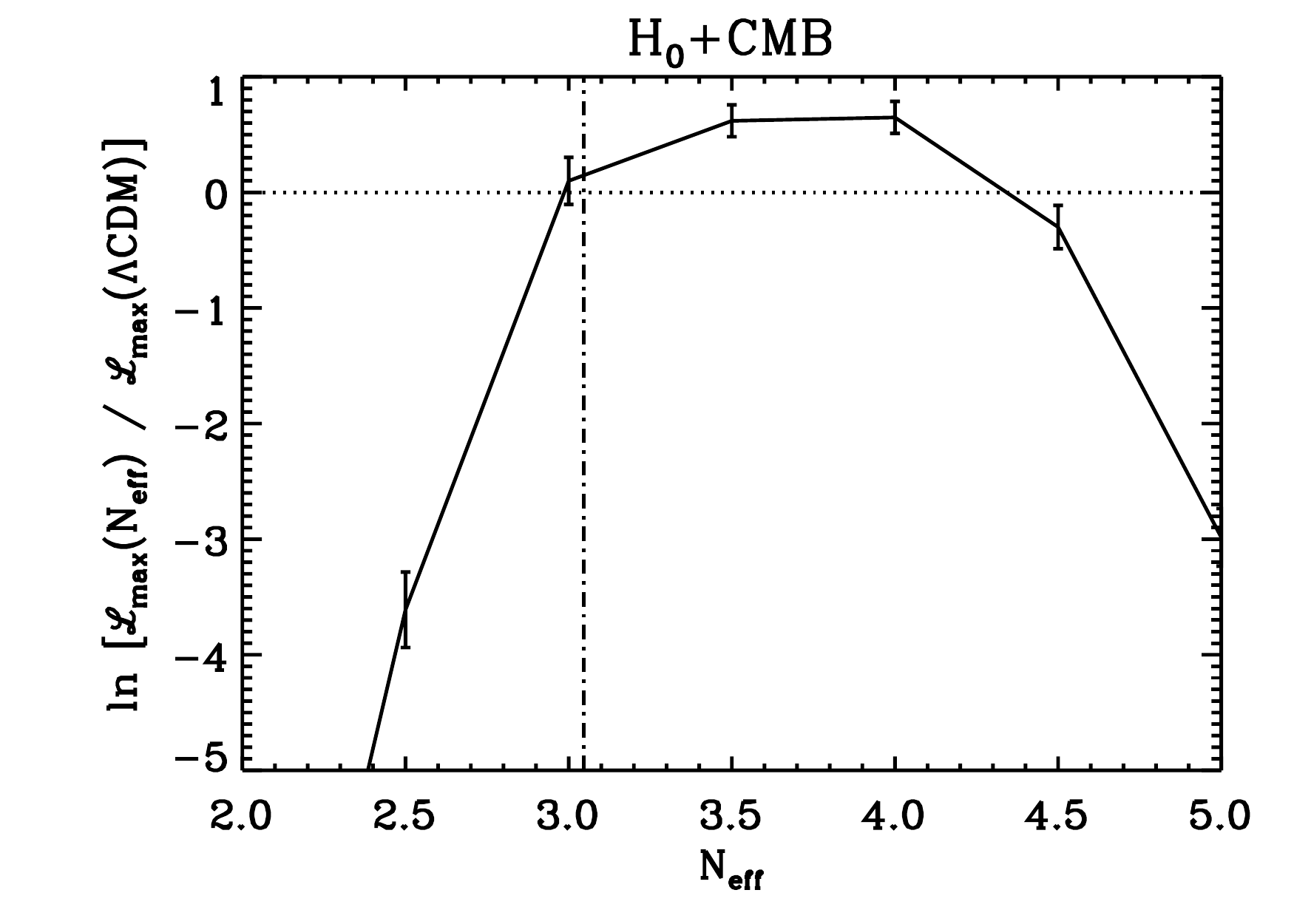}
\caption{Profile likelihood ratios comparing vanilla $\Lambda$CDM to $\Lambda$CDM with a non-standard effective number of neutrino species. Results are shown for the WMAP (top left), WMAP+SPT and WMAP+SPT+SPTLens (top right, solid and dashed lines, respectively), WMAP+SPT+SPTLens+BAO (bottom left) and WMAP+SPT+SPTLens+$\hnought$ (bottom right) dataset combinations. Estimates on the uncertainty of the PLR values are plotted; the standard value of $\neff$ is indicated by the dot-dashed line.}
\label{fig:neff_profile_likes}
\end{figure}

The PLR for $\neff$ assuming the $\Lambda$CDM$\neff$ model is plotted in Fig.~\ref{fig:neff_profile_likes}.
To generate these plots, the PLR is calculated as described in Eq.~\ref{eq:plr_proxy}, identifying $\pars$ with the standard parameters of $\Lambda$CDM and $\beta$ with $\neff$, and using bins of width $\Delta \neff = 0.5$. For illustrative purposes, we have chosen to normalize with respect to the maximum likelihood found for pure $\Lambda$CDM. Estimates of the scatter on the PLR introduced by sampling are obtained by first finding the difference between the maximum $\Lambda$CDM likelihood and the maximum $\Lambda$CDM$\neff$ likelihood in a small bin of width 0.01 centered on $\neff = 3.046$. The error on each PLR bin is then estimated by weighting this difference by the square root of the ratio of the number of samples in the PLR bin to the number of samples in the small $\neff \simeq 3.046$ bin.

For all dataset combinations, the standard value of $\neff$ is well within two-$\sigma$ of the maximum likelihood recorded: indeed, taking into account the sampling-induced scatter, the difference is on the one-$\sigma$ level for all but the SPT+WMAP datasets. This is in good agreement with the posterior parameter constraints, and indicates at most a mild preference in the data for a non-standard value of $\neff$. This preference is not large enough to push the Bayesian Evidence in favor of the extended model.


\section{Conclusions}
\label{sec:conclusions}

Recent cosmological and particle physics results have suggested that the current model of the neutrino should be extended to include additional species and/or mass. In this work, we have applied Bayesian model selection to determine whether current cosmological datasets truly favor models with extended neutrino physics over the standard picture. The datasets considered include CMB temperature and polarization power spectra, measurements of the BAO scale and the Hubble constant, as well as an integrated measurement of the growth of structure in the form of the CMB lensing power spectrum. The model selection results are compared to traditional (parameter estimation) and prior-independent (profile likelihood ratio) methods to understand the current findings and explore the prior-dependence of our conclusions.

Although parameter estimates for the effective number of neutrino species, $\neff$, are higher than the standard value, our Bayesian model selection results show that current cosmological data favor the standard cosmological model over one with additional neutrino species. The high estimate of $\neff$ is likely due to marginalization over a degeneracy with cosmological parameters~\cite{GonzalezMorales:2011ty}, a finding corroborated by the profile likelihood ratio. Similarly for the sum of neutrino masses, $\mnu$, we find that the current cosmological data prefer $\Lambda$CDM over $\Lambda$CDM with massive neutrinos. We also note that the SPT CMB lensing data are not precise enough to strongly impact our findings; however, the increased sky coverage provided by {\em Planck} -- publicly released after this work was submitted for publication -- now allows CMB lensing data to play a critical role in defining our model of the Universe~\cite{Planck_Params:2013}. An analysis including the newly released {\em Planck} data will be presented in a forthcoming paper~\cite{Feeney_etal:2013}.


\bibliographystyle{JHEP}
\bibliography{n_eff}
\vspace*{1cm}

\acknowledgments{SMF is supported by STFC. HVP is supported by STFC, the Leverhulme Trust, and the European Research Council under the European Community's Seventh Framework Programme (FP7/2007-2013) / ERC grant agreement no 306478-CosmicDawn. LV is supported by FP7-IDEAS-Phys.LSS 240117 and thanks the Galileo Galilei Institute for Theoretical Physics for their hospitality during the completion of this work.  We acknowledge the use of the Legacy Archive for Microwave Background Data Analysis (LAMBDA). Support for LAMBDA is provided by the NASA Office of Space Science. We thank Alex van Engelen for communications regarding the SPT lensing likelihood, and Ofer Lahav for useful conversations.}


\section*{SPT Lensing Likelihood}

The SPT lensing likelihood is a Gaussian likelihood based upon the lensing potential power spectrum, $C_L^\Phi = L^4 C_L^{\phi\phi}$, following the {\tt camb} naming convention. Recall that $L$ is the magnitude of the sum of two CMB modes in harmonic space: $L = | {\bf l}_1 + {\bf l}_2 |$. SPT measures the power spectrum in bands, $B$, of width $\Delta L = 100$, centered on $L = \{ 150,\,250\,\ldots,\,1450\}$. The log-likelihood for a model $M$ is therefore
\begin{equation}
\ln \prob(\data | M) = -\frac{1}{2} \sum_{B,B^\prime} \left[ \hat{C}_B^{\Phi} - C_B^{\Phi}(M) \right] {\bf C}_{BB^\prime}^{-1}(M) \left[ \hat{C}_{B^\prime}^{\Phi} - C_{B^\prime}^{\Phi}(M) \right]^T + \ln | {\bf C}(M) | - \ln | {\bf C}^{\rm fid} |,
\end{equation}
where $\hat{C}_B^{\Phi}$ and $C_B^{\Phi}(M)$ are the measured and predicted powers in band $B$, respectively, ${\bf C}$ is the bandpower covariance matrix, and we have chosen to normalize by subtracting off the determinant of the covariance matrix assuming a fiducial $\Lambda$CDM cosmology.

As described in Ref.~\cite{vanEngelen:2012va}, the covariance matrix is constructed from four components:
\begin{itemize}
\item the raw bandpower covariance, $\bf{C}^{\rm raw}$, estimated from 2000 lensed CMB simulations of one of the SPT fields using a fiducial $\Lambda$CDM cosmology;
\item the lensing-potential bandpowers, $C_{B{\rm ,\, fid}}^{\Phi}$, of this fiducial cosmology;
\item the so-called zeroth-order bias, $N_B^{(0)}$, sourced by confusion between the effects of lensing and uncertainty in the primordial CMB signal imprinted at last scattering, and estimated by performing the lensing reconstruction on {\em unlensed} CMB simulations, and;
\item additional uncertainty, $\bf{C}^{\rm cal}$, induced by the SPT temperature calibration: all four SPT fields are rescaled (with some small error) such that the temperature power spectrum in the range $1200 \le \ell \le 3000$ matches that of the fiducial $\Lambda$CDM cosmology.
\end{itemize}
When sampling different cosmologies, the diagonal of the raw covariance matrix is first rescaled by a factor of $\left[ (C_B^{\Phi}(M) + N_B^{(0)}) / (C_{B{\rm ,\, fid}}^{\Phi} + N_B^{(0)}) \right]^2$ to account for changes in sample variance. The resulting matrix is then added to the calibration-induced uncertainty to produce ${\bf C}(M)$.

The Fortran version of the SPT lensing likelihood used in this work can be downloaded from~\url{http://zuserver2.star.ucl.ac.uk/~smf/spt_lensing_likelihood.tar.gz}.

\end{document}